# Self-consistent Force Scheme in the Discrete Boltzmann Equation

Xuhui Li, Xiaowen Shan[*]

*Department of Mechanics and Aerospace Engineering,*
*Southern University of Science and Technology, Shenzhen, Guangdong 518055, China*

lixh6@sustc.edu.cn, shanxw@sustc.edu.cn

**Abstract**

In the work of N. Martys et al. [Nicos S. Martys, Xiaowen Shan, Hudong Chen, Phys. Rev. E, Vol. 58, Num.5, 1998 ], a self-consistent force term to any order in the Boltzmann-BKG equation is derived by the Hermite basis with raw velocity. As an extension, in the present work, the force term is expanded by the Hermite basis with the relative velocity in the comoving coordinate and the Hermite basis with the relative velocity scaled by the local temperature. It is found that the force scheme proposed by He et al. [Xiaoyi He, Xiaowen Shan, Gary D. Doolen, Phys. Rev. E, Vol. 57, Num.1,1998] can be derived by the Hermite basis with the relative velocity. Furthermore, another new force scheme in which the velocity is scaled by the local temperature is obtained.

Different from the conventional Navier-Stokes-Fourier (NSF) solver, the lattice Boltzmann method is a mesoscopic approach to describe the fluid system with the collision and stream operation of the particle distribution function (PDF). The first several low-order moments of the PDF are identical to the macroscopic variables of the fluid system. The connection of the lattice Boltzmann equation and the NSF equation can be established by the Chapman-Enskog expansion [1] in the regime of small Knudsen number. However, unlike the treatment of the source term for the body force in the NSF solver, the incorporation of the body force in the lattice Boltzmann equation is not so direct. Historically, some significant progresses have been made by N. Martys et al. [2], He et al. [3], Guo et al.[4] and Shan et al.[5]. In this work, this issue is revisited. Following the fashion adopted by N. Martys et al. [2] and Shan et al.[5], an extension work is conducted in the present work. The external force is systematically expanded by three different Hermite basis with raw velocity, relative velocity and relative velocity scaled by the local temperature, which are denoted as raw moment

(RM), central moment (CM) and relative central moment (RCM) Hermite basis. Some inherent connections are revealed by this systematical Hermite representations.

To be a start, the concise description of the Boltzmann-BGK equation is stated firstly. The Boltzmann-BGK equation [6] mainly describes the evolution of the statistical fictitious particle distribution function in the velocity space and physical space. The PDF $f(\mathbf{x},\boldsymbol{\xi},t)$ satisfies the following equation

$$\frac{\partial f}{\partial t}+\boldsymbol{\xi}\cdot\nabla f+\mathbf{g}\cdot\nabla_{\boldsymbol{\xi}}f=-\frac{1}{\tau}(f-f^{eq}) \ , \tag{1}$$

in which $f(\mathbf{x},\boldsymbol{\xi},t)$ represents the probability of the fictitious particles with specific velocity $\boldsymbol{\xi}$ in the physical location $\mathbf{x}$ at the time $t$. It can be regarded that the transport of the PDF at the left-hand side of Eq. (1) is balanced by the collision at the right-hand side. If the characteristic speed is chosen as $c_s=\sqrt{RT_0/m_0}$ ($R$ is the Boltzmann constant, $T_0$ is the reference temperature and $m_0$ is the unit mass of the gas molecules), the Maxwell-Boltzmann equilibrium PDF can be written as

$$f^{eq}(\mathbf{x},\boldsymbol{\xi},t)=\frac{\rho}{(2\pi\theta)^{D/2}}\exp[-\frac{(\boldsymbol{\xi}-\mathbf{u})^2}{2\theta}] \ , \tag{2}$$

where $\theta$ is the dimensionless temperature. If $\rho_0$ is set as the reference density, then the dimensionless distribution function can be defined as $\overline{f}=fc_s^D/\rho_0$. For simplicity, the bar notation is omitted hereafter.

Because of the great property of orthogonalization for the Hermite polynomials, they have been applied in the forerunner work by Grad [7]. In the present work, three different Hermite basis $H(\boldsymbol{\xi})$, $H(\boldsymbol{\xi}-\mathbf{u})$ and $H((\boldsymbol{\xi}-\mathbf{u})/\sqrt{\theta})$ will be considered to expand the distribution function and the force term. The mathematical derivation of the interrelations of these three Hermite basis are elaborated our previous work [8,9]. The Hermite expansion of the PDF can be regarded as a series decomposition of the distribution function to different basis. The RM Hermite decompositions are as follows

$$f(\mathbf{x},\boldsymbol{\xi},t)=\omega(\boldsymbol{\zeta})\sum_{n=0}^{\infty}\frac{1}{n!}\mathbf{q}^{(n)}:H^{(n)}(\boldsymbol{\zeta}) \ . \tag{3}$$

In the right-hand side of Eq. (3), the symbol ":" stands for the full contraction of the two n-rank tensors. The Hermite basis is defined as

$$H^{(n)}(\zeta) = \frac{(-1)^n \nabla_\zeta^n \omega(\zeta)}{\omega(\zeta)} , \qquad (4)$$

in which the weighted function associated with the Hermite polynomials is

$$\omega(\zeta) = \frac{1}{(2\pi)^{D/2}} \exp(-\zeta^2 / 2) . \qquad (5)$$

where $\zeta^2 = \zeta \cdot \zeta$ and $\zeta$ can be chosen as $\xi, \mathbf{c}, \boldsymbol{\eta}$ with $\mathbf{c} = \xi - \mathbf{u}$ and $\boldsymbol{\eta} = \mathbf{c}/\sqrt{\theta}$.

Taking the property of orthogonalization for the Hermite basis into account, the corresponding Hermite coefficients can be evaluated as

$$\mathbf{a}^{(n)} = \int f H^{(n)}(\xi) d\xi, \quad \mathbf{b}^{(n)} = \int f H^{(n)}(\mathbf{c}) d\mathbf{c}, \quad \mathbf{d}^{(n)} = \int f H^{(n)}(\boldsymbol{\eta}) d\boldsymbol{\eta}. \qquad (6)$$

Substitute Eq. (4) into Eq. (3), the following form of the distribution function can be obtained

$$f(\mathbf{x}, \xi, t) = \sum_{n=0}^{\infty} \frac{(-1)^n}{n!} \mathbf{q}^{(n)} : \nabla_\zeta^n \omega(\zeta), \qquad (7)$$

with $\mathbf{q}^{(n)} = \mathbf{a}^{(n)}, \mathbf{b}^{(n)}, \mathbf{d}^{(n)}$.

The derivative of the distribution function relative to different variables can be evaluated as

$$\nabla_\zeta f = \sum_{n=0}^{\infty} \frac{(-1)^n}{n!} \mathbf{q}^{(n)} : \nabla_\zeta^{n+1} \omega(\zeta) = -\omega(\zeta) \sum_{n=1}^{\infty} \frac{1}{n!} n \mathbf{q}^{(n)} : H^{(n)}(\zeta) . \qquad (8)$$

Thus, the force term $-\mathbf{g} \cdot \nabla_\zeta f$ in Eq. (1) can be expanded in the Hermite spectrum space of finite order:

$$F(\xi) \approx \omega(\xi) \sum_{n=1}^{N} \frac{1}{n!} n \mathbf{g} (\mathbf{a}_0^{(n-1)} + \mathbf{a}_1^{(n-1)}) : H^{(n)}(\xi), \qquad (9)$$

$$F(\xi) \approx \omega(\mathbf{c}) \sum_{n=1}^{N} \frac{1}{n!} n \mathbf{g} (\mathbf{b}_0^{(n-1)} + \mathbf{b}_1^{(n-1)}) : H^{(n)}(\mathbf{c}), \qquad (10)$$

$$F(\xi) \approx \omega(\boldsymbol{\eta}) \sum_{n=1}^{N} \frac{1}{n!} n \mathbf{g} (\mathbf{d}_0^{(n-1)} + \mathbf{d}_1^{(n-1)}) : H^{(n)}(\boldsymbol{\eta}) , \qquad (11)$$

in which the subscript 0 and 1 denotes the Hermite coefficients for the equilibrium and non-equilibrium parts. The details of these coefficients can be found in our previous work [8,9].

If the above three formula are truncated to $N=4$, then the force terms can be written as the following explicit forms

$$F(\boldsymbol{\xi}) \approx \omega(\boldsymbol{\xi})\rho\{\underbrace{\mathbf{g}\cdot\boldsymbol{\xi}}_{1st}+\underbrace{(\mathbf{g}\cdot\boldsymbol{\xi})(\boldsymbol{\xi}\cdot\mathbf{u})-\mathbf{g}\cdot\mathbf{u}}_{2nd}+\underbrace{\frac{1}{2\rho}\mathbf{g}[\rho(\mathbf{u}^2+(\theta-1)\boldsymbol{\delta})+\mathbf{a}_1^{(2)}]:H^{(3)}(\boldsymbol{\xi})}_{3rd}$$

$$+\underbrace{\frac{1}{6\rho}\mathbf{g}(\mathbf{a}_0^{(3)}+\mathbf{a}_1^{(3)}):H^{(4)}(\boldsymbol{\xi})}_{4th}\}\tag{12}$$

$$F(\boldsymbol{\xi}) \approx \omega(\boldsymbol{\xi}-\mathbf{u})\rho\{\underbrace{\mathbf{g}\cdot(\boldsymbol{\xi}-\mathbf{u})}_{1st}+\underbrace{0}_{2nd}+\underbrace{\frac{1}{2\rho}\mathbf{g}[\rho(\theta-1)\boldsymbol{\delta}+\mathbf{a}_1^{(2)}]:H^{(3)}(\boldsymbol{\xi}-\mathbf{u})}_{3rd}$$

$$+\underbrace{\frac{1}{6\rho}\mathbf{g}(\mathbf{a}_1^{(3)}-3\mathbf{u}\mathbf{a}_1^{(2)}):H^{(4)}(\boldsymbol{\xi}-\mathbf{u})}_{4th}\}\tag{13}$$

$$F(\boldsymbol{\xi}) \approx \omega(\frac{\boldsymbol{\xi}-\mathbf{u}}{\sqrt{\theta}})\rho\{\underbrace{\mathbf{g}\cdot(\frac{1}{\sqrt{\theta}})^{D+1}(\boldsymbol{\xi}-\mathbf{u})}_{1st}+\underbrace{0}_{2nd}+\underbrace{\frac{1}{2\rho}\mathbf{g}(\frac{1}{\sqrt{\theta}})^{D+2}\mathbf{a}_1^{(2)}:H^{(3)}(\frac{\boldsymbol{\xi}-\mathbf{u}}{\sqrt{\theta}})}_{3rd}$$

$$+\underbrace{\frac{1}{6\rho}\mathbf{g}(\frac{1}{\sqrt{\theta}})^{D+3}(\mathbf{a}_1^{(3)}-3\mathbf{u}\mathbf{a}_1^{(2)}):H^{(4)}(\frac{\boldsymbol{\xi}-\mathbf{u}}{\sqrt{\theta}})}_{4th}\}\tag{14}$$

Furthermore, if the first two orders of Eq. (12-14) are retained, then the following forms for the external force can be obtained

$$F(\boldsymbol{\xi}) \approx \omega(\boldsymbol{\xi})\rho[\mathbf{g}\cdot\boldsymbol{\xi}+(\mathbf{g}\cdot\boldsymbol{\xi})(\boldsymbol{\xi}\cdot\mathbf{u})-\mathbf{g}\cdot\mathbf{u}]\tag{15}$$

$$F(\boldsymbol{\xi}) \approx \omega(\boldsymbol{\xi}-\mathbf{u})\rho\mathbf{g}\cdot(\boldsymbol{\xi}-\mathbf{u})=\mathbf{g}\cdot(\boldsymbol{\xi}-\mathbf{u})f^{(eq)}=-\mathbf{g}\cdot\nabla_{\mathbf{c}}f^{(eq)}\tag{16}$$

$$F(\boldsymbol{\xi}) \approx \omega(\boldsymbol{\eta})(\frac{1}{\sqrt{\theta}})^D\rho\mathbf{g}\cdot\boldsymbol{\eta}=\mathbf{g}\cdot\frac{\boldsymbol{\xi}-\mathbf{u}}{\sqrt{\theta}}f^{(eq)}=-\mathbf{g}\cdot\nabla_{\boldsymbol{\eta}}f^{(eq)}\tag{17}$$

In the above derivation, the Maxwell-Boltzmann equilibrium distribution function (Eq. 2) is used and the uniform global reference temperature is applied in Eq. (16). To clarify the relationship of the force terms of Eq. (12) and Eq. (13), the first two orders and partial third-order are retained for Eq. (12) while only the first order of Eq. (13) is retained.

For Eq. (12), the first two orders and the third-order term excluding the terms concerning temperature and non-equilibrium part can be written as

$$F(\xi) \approx \omega(\xi)\rho\{\underbrace{\mathbf{g}\cdot\xi}_{1st} + \underbrace{(\mathbf{g}\cdot\xi)(\xi\cdot\mathbf{u}) - \mathbf{g}\cdot\mathbf{u}}_{2nd} + \underbrace{\frac{1}{2\rho}\mathbf{g}(\rho\mathbf{u}^2):H^{(3)}(\xi)}_{3rd}\}$$

$$= \omega(\xi)\rho\{\mathbf{g}\cdot\xi + (\mathbf{g}\cdot\xi)(\xi\cdot\mathbf{u}) - \mathbf{g}\cdot\mathbf{u} - (\mathbf{g}\cdot\mathbf{u})(\xi\cdot\mathbf{u}) + \frac{1}{2}(\mathbf{g}\cdot\xi)[(\xi\cdot\mathbf{u})^2 - u^2]\} \quad (18)$$

For Eq. (13), only the first term is retained but different terms are retained in the Taylor expansion of the exponential function.

$$F(\xi) \approx \omega(\xi - \mathbf{u})\rho\mathbf{g}\cdot(\xi - \mathbf{u})$$

$$= \omega(\xi)\rho\mathbf{g}\cdot(\xi - \mathbf{u})\exp[\xi\cdot\mathbf{u} - \frac{1}{2}u^2]$$

$$= \omega(\xi)\rho\mathbf{g}\cdot(\xi - \mathbf{u})[1 + \xi\cdot\mathbf{u} - \frac{1}{2}u^2 + O(u^3)]$$

$$= \omega(\xi)\rho\{\mathbf{g}\cdot\xi - \mathbf{g}\cdot\mathbf{u} + (\mathbf{g}\cdot\xi)(\xi\cdot\mathbf{u}) - (\mathbf{g}\cdot\mathbf{u})(\xi\cdot\mathbf{u}) + O(u^3)\} \quad (19)$$

In the above Taylor expansion of the exponential function, only zero and first orders are retained. In the following expression, one of the second-order terms is retained in the Taylor expansion.

$$F(\xi) \approx \omega(\xi - \mathbf{u})\rho\mathbf{g}\cdot(\xi - \mathbf{u}) = \omega(\xi)\rho\mathbf{g}\cdot(\xi - \mathbf{u})\exp[\xi\cdot\mathbf{u} - \frac{1}{2}u^2]$$

$$= \omega(\xi)\rho\mathbf{g}\cdot(\xi - \mathbf{u})[1 + \xi\cdot\mathbf{u} - \frac{1}{2}u^2 + \frac{1}{2}(\xi\cdot\mathbf{u})^2 + O(u^3)]$$

$$= \omega(\xi)\rho\{\mathbf{g}\cdot\xi - \mathbf{g}\cdot\mathbf{u} + (\mathbf{g}\cdot\xi)(\xi\cdot\mathbf{u}) - (\mathbf{g}\cdot\mathbf{u})(\xi\cdot\mathbf{u}) + \frac{1}{2}(\mathbf{g}\cdot\xi)[(\xi\cdot\mathbf{u})^2 - u^2]$$

$$\underbrace{-\frac{1}{2}(\mathbf{g}\cdot\mathbf{u})[(\xi\cdot\mathbf{u})^2 - u^2]}_{O(u^3)} + O(u^3)\} \quad (20)$$

The discrete forms of the above force terms can be obtained as the approach in [5] as $F_\alpha = \frac{\omega_\alpha}{\omega(\xi_\alpha)}F(\xi_\alpha)$. The discrete lattice effect could be derived with the approach proposed by He et al. [3].

Some comments should be addressed here: 1) If the first two orders are retained, Eq. (12) is the continuum counterpart of the force scheme proposed by Guo et al. [4] before considering the discrete lattice effect, which was analyzed by Shan et al. [5]; 2) In the Hermite expansion with relative velocity in the comoving coordinate, if the first order term is retained and the uniform global reference temperature is applied, Eq. (16) is definitely the force scheme proposed by He et al. [3] within the Navier-Stokes level, which was criticized to be an approximation form without serious proof [2]; 3) With the asymptotic expansion of the exponential function, the first two orders of the force scheme in Eq. (18) can be recovered by retaining the first-order expansion in Eq. (19); Third-order term in Eq. (18) can be recovered by retaining one of the second-order term in the Taylor expansion in Eq. (20). This may explain the relationship between the force schemes of Martys et al. [2] and Guo et al. [4] in contrast with that of He et al. [3]. In general, the force scheme proposed by Guo et al. [4] can be formulated by the raw moment Hermite basis while the force scheme proposed by He et al. [3] could be formulated by the central moment Hermite basis. To be a new form, Eq. (17) is an improved version of the force scheme by He et al. [3], in which the relative velocity is scaled by the local temperature, instead of the uniform global reference temperature. This new finding may reveal the intrinsic connection between the momentum equation and the energy equation when the conservation laws are resolved in the fully coupled way with only one distribution function in the high-order lattice Boltzmann model [5]. The new force scheme is speculated to be more reasonable in simulating the thermal flow with body force.

In summary, we have systematically derived the external force term in the continuum Boltzmann BGK equation by three different Hermite basis with the raw velocity, relative velocity and the relative velocity scaled by the local temperature. It is found that the first Hermite representation is coincided with the force scheme of Guo et al. [4] while the force scheme developed by He et al. [3] can be derived by the second Hermite representation. Furthermore, a refined force scheme in which the velocity is scaled by the local temperature can be obtained by the third Hermite representation. In a word, we can easily derive the forcing term in a self-consistent fashion to any order with three Hermite basis in discrete Boltzmann models.

**Acknowledgements**

This work was supported by the National Science Foundation of China Grants 91741101 and 91752204. X.L. acknowledges the financial support from the funding of the SUSTC Presidential Postdoctoral Fellowship.